# Online Static Security Assessment of Power Systems based on Lasso Algorithm

**Yahui Li [1], Yang Li [1,\*] and Yuanyuan Sun [2]**

[1] School of Electrical Engineering, Northeast Electric Power University, Jilin 132012, China; liyh1993@gmail.com
[2] School of Electrical Engineering, Shandong University, Jinan 250061, China; sunyy@sdu.edu.cn
\* Correspondence: liyang@neepu.edu.cn; Tel.: +86-159-4796-6691

**Featured Application: The proposed approach may find potential applications in fast evaluation of the security level of power systems and identification of the most severe cases from the entire contingency list.**

**Abstract:** As one important means of ensuring secure operation in a power system, the contingency selection and ranking methods need to be more rapid and accurate. A novel method-based least absolute shrinkage and selection operator (Lasso) algorithm is proposed in this paper to apply to online static security assessment (OSSA). The assessment is based on a security index, which is applied to select and screen contingencies. Firstly, the multi-step adaptive Lasso (MSA-Lasso) regression algorithm is introduced based on the regression algorithm, whose predictive performance has an advantage. Then, an OSSA module is proposed to evaluate and select contingencies in different load conditions. In addition, the Lasso algorithm is employed to predict the security index of each power system operation state with the consideration of bus voltages and power flows, according to Newton–Raphson load flow (NRLF) analysis in post-contingency states. Finally, the numerical results of applying the proposed approach to the IEEE 14-bus, 118-bus, and 300-bus test systems demonstrate the accuracy and rapidity of OSSA.

## 1. Introduction

### 1.1. Background and Motivation

In power system operations, the security of a system has always been an important issue, which is related to the ability to continue normal operation in post-contingency conditions [1]. Power system security assessment is an effective tool for checking the security of power systems, which aims to determine whether, and to what extent, a power system is reasonably safe from serious interference to its operation [2,3]. Power system security assessment includes static security assessment (SSA) and dynamic security assessment (DSA). Static security assessment is concerned with factors related to the insecurity situation [4,5], such as overload, overvoltage, and so on, via the load flow calculation of the power system in post-contingency conditions. Meanwhile, dynamic security assessment mainly analyzes the transient stability of the power system after the fault of the power system according to the real-time data [6–8]. To predict the transient stability status by machine learning algorithms in the post-fault condition, the real-time data are respectively obtained from phasor measurement units (PMUs) in [6] and the simulation results in [7,8]. This article focuses on static security assessment problems, especially for online applications. The power system static security assessment can be divided twofold, into system monitoring and contingency analysis [9]. The system monitoring can provide the operating conditions of the power system to the operators, and the updated information contains bus voltages, currents, power flows, and so on. The contingency analysis is employed to evaluate an outage event in the power system. Once the system is in an insecure operation condition, the corresponding security controls such as preventive or corrective control actions will be started up immediately to ensure the insecure outage event back to the security condition [10,11]. Although two major factors are involved in static security assessment, contingency analysis, which includes contingency screening and ranking, plays a more critical role [9].

*1.2. Literature Review*

Studies of power system static security assessment have been carried out for a few years [12–14] that have focused on contingency screening or ranking. In order to select and screen contingencies in static security assessment, the indices representing voltages deviation and line overloads are employed in [12,13]. While in Krishnan et al. [14], a database generation method is proposed for the critical contingencies that have been screened. Thus, as a commonly used method, the severity of the power system in post-contingency conditions is evaluated by the performance indices, which are computed by the variables of power systems [15]. In Sunitha et al. [16], a single composite security index, which is obtained by Newton–Raphson load flow (NRLF) analysis, is proposed for defining the security, which provides a reference for the proposed method in this paper. The composite security index divides the power system conditions into three categories: the secure state with the index value of '0'; the alarm state with the index value between '0' to '1'; and the index value greater than '1', which represents an insecure state. The composite security index considers the violations of bus voltages and line flows, and avoids the selecting of index weights [16]. By adopting the security index in the proposal, the time consumed to evaluate severity could be reduced. However, in the research on static security assessment, the literature has an excessive focus on the accuracy and celerity [15,16]. Furthermore, many of these studies ignore the regulation ability of the power system by employing adjustable devices, such as the transformers and reactive power compensation devices in the power system [17–19]. In Li et al. [17], Dall'Anese et.al [18] and Li et al. [19], the power system demonstrated improved economic and environmental benefits by optimizing the variables of adjustable devices. Therefore, on the premise of ensuring fast and accurate screen contingencies, the effects of the adjustable devices are considered in this paper.

Due to the large computing scale of NRLF analysis for obtaining the composite security index in the contingency analysis [20], the static security assessment of a power system is a heavy task. What's more, the online application of static security assessment has a higher requirement for efficiency. Thus, adopting machine learning methods is an accurate and powerful way for contingency analysis [21], whose speed and accuracy are suitable for online applications. There are three typical machine learning methods that can be used for static security assessment. One is the artificial neural network (ANN) algorithm [22−24]. In Sunitha et al. [16], the static security index is predicted by adopting the ANN algorithm for contingency screening and ranking. In Al-Masri et al. [22] and Zhou et al. [23], the ANN algorithm is applied for enhancing the security of power systems. In Varshney et al. [24], an ANN module is employed for static security assessment, considering the voltage and load flow in the power system. Although the ANN algorithm has been employed in the literature, it is inappropriate for large-scale data modules, and the internal mechanism is difficult to understand. Another is the decision tree (DT) algorithm [25,26]. In Oliveira et al. [25], the static security assessment applies machine learning techniques, which are based on decision tree algorithms, to improve the efficiency of contingency screening and ranking. In Saeh et al. [26], a modified decision tree algorithm is employed for static security assessment. Static security problems can be assessed by adopting decision tree algorithms; however, it is possible to over-fit or trap decision tree algorithms into local minimum points [27]. In addition, the support vector machine (SVM) algorithm is another machine learning method for static security assessment [28,29]. In Kalyani et.al [28] and Kalyani et al. [29], considering line overload and voltage deviation indices in the power system, SVM algorithms are employed for evaluating static security. Nevertheless, the SVM algorithm is difficult to implement if the scale of the training sets is too large. Therefore, according to the above analyses, the prediction of the above three algorithms may be affected when static security assessment have requirements for the time-solving and calculation scales.

Studies have shown that the least absolute shrinkage and selection operator (Lasso) algorithm has stronger generalization ability and predictive performance than other machine learning methods [30–33], such as the three algorithms that have been mentioned. The Lasso algorithm has been successfully utilized in many areas of power systems, such as, the state estimation of a power system in [30], image processing in [31], and telecommunications in [32]. Lasso is also used for the prediction of the transient stability boundary in [33], while a systematic approach for categorizing the model parameters based on Lasso is presented in [34]. Compared to other learning algorithms, Lasso is more accurate and has the ability for the automatic feature selection to yield a low-dimensional solution [33]. As the observation of online static security assessment (OSSA) is essentially a high-dimensional problem, the Lasso algorithm is suitable for OSSA. However, the application of the Lasso algorithm in OSSA is finite. The proposal provides an idea for the utilization of the Lasso algorithm in the area of OSSA.

*1.3. Contributions*

The purpose of this paper is to develop an OSSA module based on the Lasso algorithm. In order to reduce the computation scale of real-time static security assessment, the proposed method is applied to fast and accurate contingency screening and ranking. The proposal composes the following procedure. (1) The security index $PI_c$ is used for evaluating the security of power system operating states, and the conditions can be distinguished into secure, alarmed, and insecure states. (2) The Lasso module is trained by adopting the variables in the power system, the status of all of the lines, and the corresponding security index. (3) Once the Lasso module is trained, the proposal can predict the security index based on the variables from the power system, and the operating state also can be obtained according to the value of $PI_c$.

The main contributions of this paper are as follows. (1) In order to screen and rank the severity of contingencies, an online static security assessment module is presented considering the impacts of adjustable devices, and the operating state also can be identified via the variables of the current operating point; (2) A novel method based on the multi-step adaptive Lasso (MSA-Lasso) regression algorithm is developed for predicting the security index with the consideration of bus voltages and power flows, which is employed to evaluate the operating states; (3) Load conditions varying from 50% to 150% of the base load are considered for adapting to the changeful power system states.

*1.4. Organization of This Paper*

The remaining parts of this paper are organized as follows: A brief introduction of online security assessment in the power system is given in the following section. Then, the basic principle of the Lasso algorithm is provided in Section 3. Moreover, the Lasso algorithm employed for OSSA is proposed in Section 4. Case studies and results that demonstrate the effectiveness are carried out in Section 5, and finally, Section 6 gives the conclusion.

**2. Online Static Security Assessment in Power System**

In the modern energy management system (EMS), static security assessment is an important control method [1–3]. The assessment can be implemented according to the fault set. The set may contain contingencies with a single line outage, loss of generators, load variations, and so on. Line outages and load variations are considered in this paper. Then, three operating states of power systems—insecurity, alarmed, and security states—can be identified via static security assessment. The operators can take appropriate measures for each state based on the severities of faults.

In conventional practices, security assessment is solved by repeatedly computing the power flow for all of the pre-defined outage conditions. However, the size of the fault set is generally large, and changes with the load conditions of the power system. Static security assessment will consume a large amount of time, and it is difficult to achieve in online application. Hence, the rapid and accuracy of screening and ranking contingencies is a significant issue. For this purpose, the composite security index *PIc* is employed for evaluation, and the Lasso algorithm is applied instead of a power flow calculation to obtain the security index *PIc* in the paper.

## 3. Principles of Multi-step Adaptive Lasso Regression Algorithm (MSA-Lasso)

The principles of Lasso are briefly introduced in this section. As a regression learning algorithm, Lasso is superior to others [30–33]. The basic idea of Lasso is to minimize the residual sum of the square, under the condition that the sum of the absolute value of the regression coefficient is less than a constant. Then, some regression coefficients are produced, and they are strictly equal to 0. For a better presentation of the Lasso regression algorithm, the algorithm model is firstly introduced. Then, the regression algorithm is displayed. In addition, the Lasso regression algorithm is compared with the Ridge Regression algorithm. Finally, the MSA-Lasso regression algorithm is discussed.

### *3.1. Algorithm Model*

As a supervised machine learning algorithm, the original training set of this problem is:

$$TS = \left[ \left( x_1^{(0)}, y_1^{(0)} \right), \cdots \quad \cdots \right] \tag{1}$$

where $x_i^{(0)}$ is the $i$th original observation vector that is constructed by controlled variables in this paper, $y_i^{(0)}$ is the $i$th original response, which is the security index in this paper, and $S$ is the size of the training set. Generally, the original observation data $x_i^{(0)}$ needs to be standardized, while $y_i^{(0)}$ needs to be zero-centered; the post-treatment data $x_i$ and $y_i$ are defined as:

$$\begin{aligned} x_i &= \frac{x_i^{(0)} - \mu_x}{std_x} \\ y_i &= y_i^{(0)} - \mu_y \end{aligned} \tag{2}$$

where $\mu_x$ and $\mu_y$ are respectively the means of $x_i^{(0)}$ and $y_i^{(0)}$, and $std_x$ denotes the standard deviation of $x_i^{(0)}$. After the process, it can be ensured that $\sum_{i=1}^{S} x_i = 0$, $\sum_{i=1}^{S} x_i^2 = 1$, and $\sum_{i=1}^{S} y_i = 0$.

Taking two-dimensional data as an example, the effect after the treatment of centralization and standardization are shown in Figure 1. From the figure, it can be clearly seen that centralization moves the center of the data to the origin point, and standardization ensures that the scale of the data have the same value in each dimension.

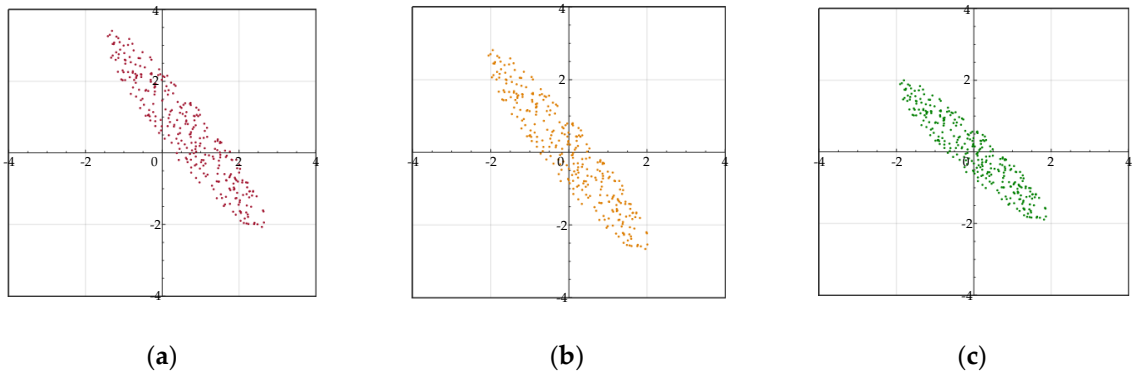

(**a**) (**b**) (**c**)

**Figure 1.** The effect after the treatment of data. (**a**) Original data; (**b**) Centralized data; (**c**) Standardized data.

The regression algorithm is based on the following simplified linear model [33]:

$$\mathbf{Y} = \mathbf{X}\sigma + \varepsilon \tag{3}$$

where $\mathbf{Y}$ is the vector consisting of responses, and $\mathbf{Y} = [y_1, \cdots \quad \cdots ]$; $\mathbf{X}$ is the $S \times D$ design observation matrix, which is formed by observation of $x_1, \cdots \quad \cdots$, and the observation $x_i$, which includes $D$ variables, has the corresponding response $y_i$; $\sigma = [\sigma_1, \cdots \quad \cdots ]$ is the vector of regression coefficients; and $\varepsilon = [\varepsilon_1, \cdots \quad \cdots ]$ is the vector of random variables produced by the innovation process. For obtaining regression coefficients $\sigma$, the linear model can be solved via the least square (LS) method. The cost function $f(\sigma)$ of the optimization problem is described as:

$$\min f_{ls}(\sigma) = \frac{1}{S} \|\mathbf{Y} - \mathbf{X}\sigma\|^2 = \frac{1}{S} \sum_{i=1}^{S} (y_i - x_i \sigma)^2 \tag{4}$$

The regression coefficients can be obtained, and the expression of estimates $\hat{\sigma}$ is:

$$\hat{\sigma} = (\mathbf{X}^T \mathbf{X})^{-1} \mathbf{X}^T \mathbf{Y} \tag{5}$$

If the estimates $\hat{\sigma}$ exist, the only condition is that $\mathbf{X}$ is a full matrix. However, even though the condition is satisfied, large relevance features of observation data will cause big errors. Therefore, the regression algorithm is proposed to improve the situation.

*3.2. Regression Algorithm*

A penalty term $R(|\sigma_i|)$ is added to the regression algorithm, which is an improvement on the least square method. The cost function $f(\sigma)$ of the regression problem is given by:

$$\min f(\sigma) = \frac{1}{S} \|\mathbf{Y} - \mathbf{X}\sigma\|^2 + \lambda \sum_{i=1}^{S} R(|\sigma_i|) \tag{6}$$

where $\lambda \geq 0$ denotes the shrinkage tuning parameter. Equation (6) is equivalent to the following functions:

$$\min f(\sigma) = \frac{1}{S} \|\mathbf{Y} - \mathbf{X}\sigma\|^2$$
$$\text{s.t.} \sum_{i=1}^{S} R(|\sigma_i|) \leq t \tag{7}$$

where $t \geq 0$ represents the tuning parameter. If the penalty term adopts $L_1$ norm, that is $R(|\sigma_i|) = |\sigma_i|$, the following optimization issue can represent the Lasso regression algorithm, and the cost function is shown as:

$$\min f_{lasso}(\sigma) = \frac{1}{S} \|\mathbf{Y} - \mathbf{X}\sigma\|^2 + \lambda \sum_{i=1}^{S} |\sigma_i| \tag{8}$$

*3.3. Multi-step Adaptive Lasso Regression Algorithm*

In order to solve the regression problem, in the basis of the classic Lasso regression algorithm which is shown as Equation (3), the constant term of estimator $\sigma_0$ is joined in the Lasso algorithm model, and it is given by:

$$\mathbf{Y} = \sigma_0 + \mathbf{X}\sigma + \varepsilon \tag{9}$$

For solving the Lasso model, the following optimization issue needs to be solved:

$$\min f(\sigma_0, \sigma) = \frac{1}{S} \sum_{i=1}^{S} l(y_i, x_i, \sigma_0, \sigma) + \lambda \|\sigma\| \tag{10}$$

where $l(y_i, x_i, \sigma_0, \sigma)$ represents the function of likelihood contribution for observation $x_i$. Aiming at solving the shortcomings of low efficiency and high complexity in the classic adaptive Lasso regression algorithm, the MSA-Lasso regression algorithm, which is realized by repeatedly using the adaptive Lasso regression algorithm, is introduced as follows.

Firstly, the Gaussian model is employed in this paper for the linear regression issue, and the above optimization issue in Equation (10) can be written as:

$$\min f(\sigma_0, \sigma) = \frac{1}{2S} \sum_{i=1}^{S} (y_i - \sigma_0 - x_i \sigma)^2 + \lambda \|\sigma\| \tag{11}$$

Then, cyclic coordinate descent (CCD) is employed for solving the optimization, which is a fast method compared with others under the current research background [35]. The basic criterion of CCD is that each parameter is optimized and circulated until the conditions of the other parameters are fixed, and the coefficient is stable. Thus, the Lasso regression algorithm is implemented multiple times. Then, the current *k*th estimator $\tilde{c}$ can be computed as:

$$\tilde{c} \quad \left( \frac{1}{S} \sum_{i=1}^{S} x_{ik}(y_i - \tilde{c}) \right) \quad \tag{12}$$

$$s.t. \quad \tilde{c} \quad \tilde{c} \quad _{l=1, j \neq k}$$

where $\tilde{c}$ and $\tilde{c}$ $\cdots$ are the current regression coefficients. Therefore, the model of the MSA-Lasso regression algorithm can be explained.

*3.4. Parameters in This Section*

In order to a better presentation of the aforementioned symbols, the main parameters in this section are listed in Table 1 as follows.

**Table 1.** The main parameters in Section 3.

| Parameter | Meaning |
|---|---|
| $x_i^{(0)}$ | The *i*th original observation vector |
| $y_i^{(0)}$ | The *i*th original response |
| $x_i$ | The *i*th post-treatment observation vector |
| $y_i$ | The *i*th post-treatment response |
| $S$ | The size of the training set |
| $D$ | The number of variables |
| **X** | The $S \times D$ design observation matrix formed by observation vectors |
| **Y** | The response vector consisting of $S$ responses |
| $\lambda$ | The shrinkage tuning parameter |
| $\sigma$ | The vector of regression coefficients |

**4. Online Static Security Assessment Method**

The proposed OSSA method is introduced in this section, and the method contains the OSSA module and the Lasso module. The OSSA module concerns the power system operating points for screening and ranking contingencies, while the Lasso module is based on the variables of the power system and security index. By predicting the security index of the power system via the Lasso module, the system static security can be assessed online.

*4.1. Overall online static security assessment method*

According to the above analysis, the overall structure of the proposed OSSA method is shown in Figure 2. The operating points are chosen as the inputs of the OSSA module, which is represented by the active power outputs $P_G$ and reactive power outputs $Q_G$ of the generators, the active loads $P_D$ and reactive loads $Q_D$ of the load buses, and the voltage amplitudes $U$ and angles $\delta$ of the buses. In this paper, the light, normal, and heavy load conditions are considered, and the load condition randomly varies from 50% to 150% of the base load for representing different load conditions. Then, the contingency screening and ranking can be accomplished. In addition, the operating state of the corresponding operating point can be identified.

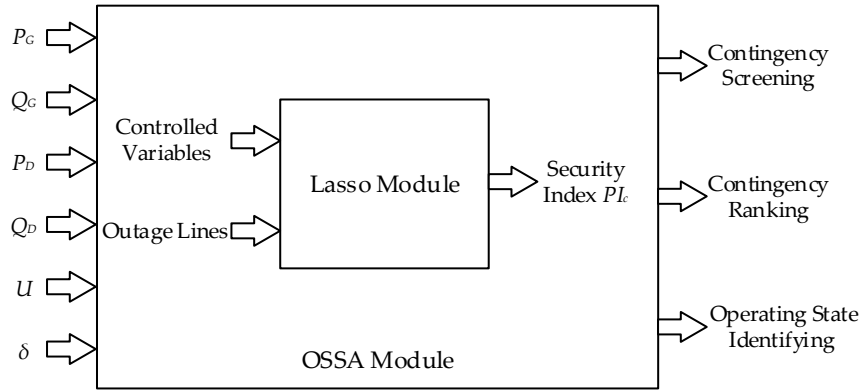

**Figure 2.** Overall structure of the online static security assessment method.

The MSA-Lasso algorithm is employed in the Lasso module, and the Lasso module is trained by taking controlled variables, which are displayed in Section 3.2 as observations, while the security index *PIc* is taken as a response introduced in Section 3.3. Based on the Lasso module, the OSSA can run fast and accurately.

*4.2. Selection of Observations Considering Adjustable Devices*

The MSA-Lasso algorithm is adopted to prediction, and the observations are divided into two parts: one part is the states of all of the lines in the power system comprising running lines, which are represented by 1, and outage lines, which are represented by 0; the other is controlled variables among the variables in an AC power system, and the controlled variables are generated randomly in their ranges.

In order to achieve better regulation and control of the power system, the impacts of the transformer and compensation equipment are taken into account in this paper. Owing to the controlled variables that vary in the process of load flow computation, they are selected as one part of the observations of the MSA-Lasso algorithm. The controlled variables and their ranges are listed as follows:

$$\begin{aligned} P_{G_i \min} \le P_{G_i} \le P_{G_i \max}, & \quad i = 1, \cdots \\ Q_{G_i \min} \le Q_{G_i} \le Q_{G_i \max}, & \quad i = 1, \cdots \\ U_{i \min} \le U_i \le U_{i \max}, & \quad i = 1, \cdots \\ T_{i \min} \le T_i \le T_{i \max}, & \quad i = 1, \cdots \\ Q_{C_i \min} \le Q_{C_i} \le Q_{C_i \max}, & \quad i = 1, \cdots \end{aligned} \quad (13)$$

where $P_{G_i}$ and $Q_{G_i}$ are respectively the active and reactive power outputs of generator *i*, $U_i$ is the magnitude of voltage in bus *i*, $T_i$ is the tap of transformer *i*, $Q_{C_i}$ is the switching capacity of reactive power compensation capacitor *i*, and $N_G$, $N_{ac}$, $N_T$ and $N_C$ are respectively the numbers of generators, buses, adjustable transformer taps, and reactive power compensation capacitor banks in the AC power system; the upper and lower limits of each variable are distinguished by the subscripts min and max.

*4.3. Selection of Responses Considering Bus Voltages and Power Flows*

After NRLF analysis, the corresponding security index *PIc* is chosen as the response of the MSA-Lasso algorithm. The index can assess the security of the power system under *N*−1 contingencies [16], and *PIc*, which considers bus voltages and power flows, is defined as:

$$PI_c = \left[ \sum_i \left(q_{U_i}^{\max}\right)^{2n} + \sum_i \left(q_{U_i}^{\min}\right)^{2n} + \sum_j \left(q_{P_j}\right)^{2n} \right]^{\frac{1}{2n}}$$

$$s.t.\ q_{U_i}^{\max} = \begin{cases} \dfrac{U_i - H_i^{\max}}{A_i^{\max} - H_i^{\max}}, & \text{if } U_i > H_i^{\max} \\ 0, & \text{otherwise} \end{cases}$$

$$q_{U_i}^{\min} = \begin{cases} \dfrac{H_i^{\min} - U_i}{H_i^{\min} - A_i^{\min}}, & \text{if } U_i < H_i^{\min} \\ 0, & \text{otherwise} \end{cases} \quad (14)$$

$$q_{P_j} = \begin{cases} \dfrac{|P_j| - P_{H_j}}{P_{H_j} - P_{A_j}}, & \text{if } |P_j| > P_{H_j}; \\ 0, & \text{otherwise} \end{cases}$$

Where *n* is the exponent, and *n*=2 in this paper; $A_i^{\max}$ and $A_i^{\min}$ are respectively the upper and lower alarm limits of $U_i$; $H_i^{\max}$ and $H_i^{\min}$ are respectively the upper and lower security limits of $U_i$; $P_{A_j}$ and $P_{H_j}$ are the upper alarm limit and security limit of the *j*th line active power flow $P_j$. In this equation, $q_{U_i}^{\max}$ and $q_{U_i}^{\min}$ are employed for evaluating the voltage limit violations, while $q_{P_j}$ can determine the extent of line overloading. In the power flow calculation, there are no limits of power flows; thus, the value of *PIc* can be very large, and the security index can better evaluate the severity of contingency caused by overloading.

By adopting the security index *PIc* as the response, the operating states are distinguished into three classes as following:

1. $PI_c = 0$, the system is in the secure state,
2. $0 < PI_c \leq 1$, the system is in the alarmed state,
3. $PI_c > 1$, the system is in the insecure state.

The contingencies can be identified as one of the three states, and the severity can be sorted by the values of security index *PIc*, where a higher value of *PIc* means a more serious contingency.

*4.4. Solution Process of Lasso Module*

According to the observations and responses of the MSA-Lasso algorithm, the solution procedures of the Lasso module are listed as follows, and the flowchart is as shown in Figure 3.

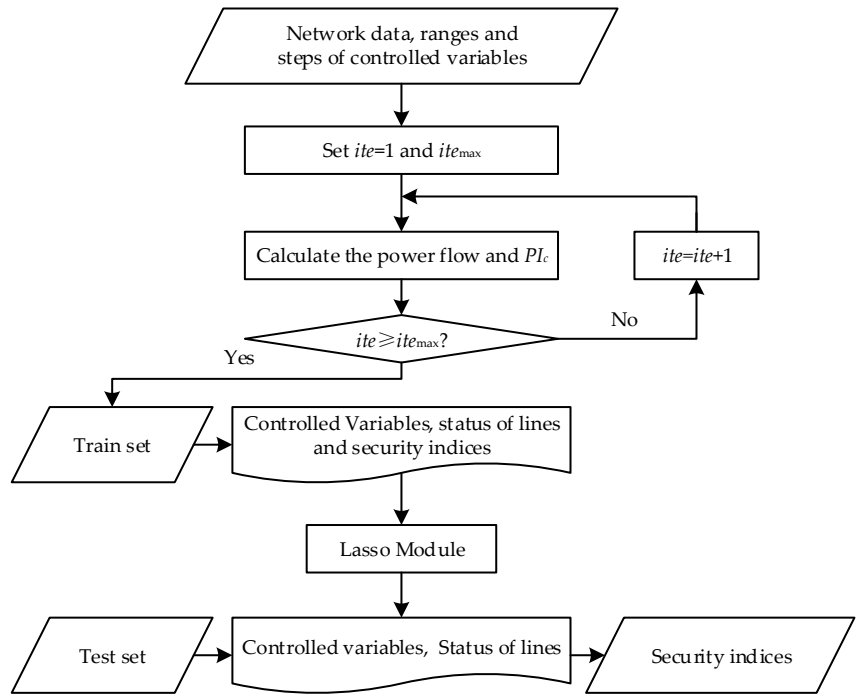

**Figure 3.** Solution process of the least absolute shrinkage and selection operator (Lasso) module.

Step 1: Input the data of the network, particularly the load conditions (i.e., light, normal, heavy load conditions). Input the ranges of controlled variables in Equation 14. In addition, the steps of the discontinuous variables, $T$ and $Q_C$, also need to be provided.

Step 2: Set the iteration number $ite=1$, and its maximum value $ite_{max}$.

Step 3: Calculate the load flow of the power system with randomly controlled variables in the condition that every line has an outage respectively; then, the corresponding response $PI_c$ is computed. In this step, the information is recorded, which includes the random controlled variables, the outage line, and the corresponding index $PI_c$.

Step 4: Judge the terminate conditions. If $ite \geq ite_{max}$, the cyclical process comes to an end, and continues to the following steps. Otherwise, come back to Step 3, and increase the iteration number by 1.

Step 5: The Lasso module is trained with conservations and corresponding responses, where each conservation is the set formed by the random controlled variables and the status of lines, and each response is the security index $PI_c$. Moreover, it should be noted that the size of Lasso training set $S$ is same as the maximum iteration number $ite_{max}$.

Step 6: The security index $PI_c$ can be predicted by adopting the Lasso module, with a known set of controlled variables and outage lines in the power system.

*4.5. Solution Using Online Static Security Assessment Method*

Based on the above overall frame of OSSA and the introduction of the Lasso module, the proposed method is discussed in this section. The online static security assessment method is presented by two parts: the OSSA module and the Lasso module. The application of the proposed method is discussed as follows.

In the OSSA module, the load condition randomly varies from 50% to 150% of the base load; thus, the load condition needs to be identified. According to the values of active loads $P_D$ and reactive loads $Q_D$, the power system data are divided into different types for representing different load conditions.

For the determined load condition, the set is randomly generated in the post-contingency condition. The set contains the observations and responses of the Lasso algorithm, and the observations are active power outputs $P_G$, reactive power outputs $Q_G$, voltages $U$, taps of transformers $T$, switching capacities of reactive power compensation capacitors $Q_C$, and the status of all of the lines that are represented by "0" and "1"; the responses are corresponding security indices $PI_c$. The number of sets is based on the number of outage lines and types of load conditions. Meanwhile, 20% of the sets are employed as test sets, and the others are applied as training sets. Then, each training set can obtain a Lasso module after training.

Once all of the Lasso modules are trained, the Lasso algorithm can be employed for predicting the security index $PI_c$ of a certain load condition with the line outage. According to the value of $PI_c$, the operating state of the power system can be identified, and the contingencies can be sorted and screened.

*4.6. Parameters in This Section*

In order to obtain a better presentation the aforementioned symbols, the main parameters in this section are listed in Table 2 as follows.

**Table 2.** The main parameters in Section 4.

| Parameter | Meaning |
|---|---|
| $P_G$ | Active power outputs of generators |
| $Q_G$ | The reactive power outputs of generators |
| $P_D$ | Active loads of load buses |
| $Q_D$ | Reactive loads of load buses |
| $U$ | Voltage amplitudes of buses |
| $\delta$ | Voltage angles of buses |
| $T$ | Taps of transformers |
| $Q_C$ | Switching capacities of reactive power compensation capacitors |
| $A_i$ | Alarm limits of $U_i$ |
| $H_i$ | Security limits of $U_i$ |
| $P_A$ | Upper alarm limit of line active power flow |
| $P_H$ | Upper security limit of line active power flow |
| $PI_c$ | Security index |

**5. Case Studies**

The IEEE 14-bus, 118-bus, and 300-bus systems are analyzed in this section. The programs of all of the algorithms are performed by a desktop computer with 3.40 GHz central processing unit (CPU) basic frequency and 4 GB memory in MATLAB 2013a. The effectiveness of the proposal is verified through the comparison of the actual value and predictive value of the index. Then, the security indicators are sorted for assessing the severities of contingencies.

In the test systems, the outages of every line are considered, except the lines that are the only line connected to the generator. The upper and lower alarm limits are ±5% of the desired bus voltage, and the upper and lower of security limits are ±7%. The alarm limit of line flow is 80% of the security limit. The observations and responses of the Lasso algorithm are listed in Table 3. In the table, $L$ represents the status of all of the lines with "0" or "1".

**Table 3.** Observations and responses of the Lasso algorithm.

| Parameter | Observation | | | | | | Response |
|---|---|---|---|---|---|---|---|
| **Variable** | $P_G$ | $Q_G$ | $U$ | $T$ | $Q_C$ | $L$ | $PI_c$ |

Moreover, under normal operating conditions, the ranges of active power output $P_G$ and reactive power output $Q_G$ are set according to the IEEE standard examples [36], the voltage $U$ ranges from 0.95 p.u. to 1.05 p.u., the transformer tap $T$ ranges from 0.9 p.u. to 1.1 p.u., and the switching capacity $Q_C$ ranges from 0 p.u. to 0.5 p.u. The steps of transformer tap $T$ and switching capacity $Q_C$ are respectively 0.125 p.u. and 0.01 p.u.

*5.1. IEEE 14-bus System*

5.1.1. Case Introduction

The IEEE 14-bus system in the literature [1,17,36] is used as a test case to examine the effectiveness of the proposed approach. This system includes five generators, 11 loads, 20 branches, three adjustable transformers, and one reactive power compensation device (connected to bus 9), as shown in Figure 4.

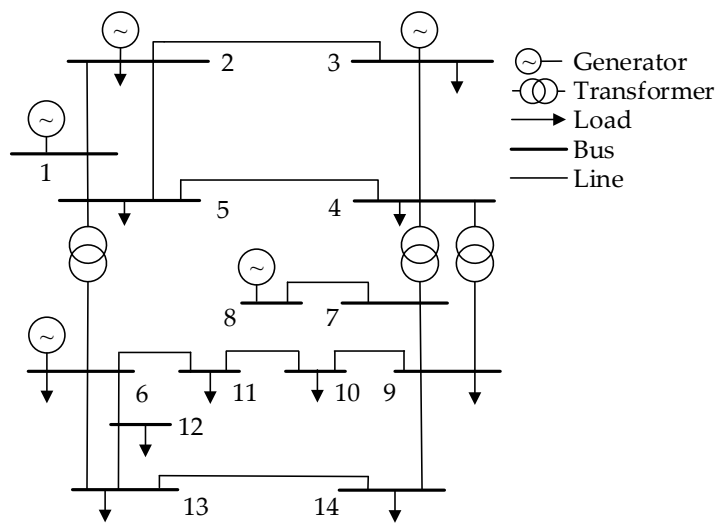

**Figure 4.** Single line diagram of the IEEE 14-bus system.

In the case of the 14-bus system, 19 lines are considered for the *N*−1 contingencies analysis. The line 7–8 (L 14) is not taken into account, because it is the only line connected to the generator. The load condition randomly varies from 50% to 150% of the base load; 950 sets are generated for the MSA-Lasso algorithm, where 20% of the sets are employed for test sets, and the others are applied as training sets [15]. It should be noted that more sets can be generated if necessary. Once the Lasso model is trained, it can be used for predicting the security indices.

5.1.2. Base Load Condition

Under the condition of *N*−1 contingencies in the normal load conditions of the IEEE 14-bus system, the values of security index $PI_c$ are computed by using the NRLF analysis results and predicted by adopting the MSA-Lasso, ANN, and SVM algorithms. One of the ANN algorithms, the back-propagation (BP) neural network, is employed in this paper, and it is a relatively commonly used ANN algorithm [9]. The parameter settings of each algorithm are listed in Table 4, and the comparison of times for contingency screening and ranking by adopting these methods are listed in Table 5.

**Table 4.** Parameter settings of the three algorithms. MSA-Lasso: multi-step adaptive Lasso; ANN: artificial neural network; SVM: support vector machine.

| Algorithm | Parameter | Value |
| --- | --- | --- |
| MSA-Lasso | Penalty parameter | 1 |
| | Number of shrinkage tuning parameters | 100 |
| ANN | Hidden layer nodes | 5 |
| | Epochs | 100 |
| | Learning rate | 0.1 |
| | Goal | 0.001 |
| SVM | Penalty parameter | 1000 |
| | Kernel parameter | 0.01 |

**Table 5.** Comparison of computation times of a 14-bus system. NRLF: Newton–Raphson load flow.

| Method | NRLF | MSA-Lasso | ANN | SVM |
|---|---|---|---|---|
| Time (s) | 0.2529 | 0.0783 | 0.1052 | 0.0927 |

It can be obviously seen that the time of the proposed module is faster than the others. So, the proposal is more available for online applications. The tested controlled variables are listed in Table 6, with the considerations of adjustable transformer taps and reactive power compensation capacitor banks.

**Table 6.** Tested controlled variables of the IEEE 14-bus system.

| Variable | Value (p.u.) | Variable | Value (p.u.) |
|---|---|---|---|
| $P_{G_1}$ | 2.3597 | $U_5$ | 1.0075 |
| $P_{G_2}$ | 0.3753 | $U_6$ | 1.0076 |
| $P_{G_3}$ | 0 | $U_7$ | 0.9607 |
| $P_{G_4}$ | 0 | $U_8$ | 1.0047 |
| $P_{G_5}$ | 0 | $U_9$ | 0.9510 |
| $Q_{G_1}$ | −0.0165 | $U_{10}$ | 0.9528 |
| $Q_{G_2}$ | 0.3743 | $U_{11}$ | 0.9760 |
| $Q_{G_3}$ | 0.0206 | $U_{12}$ | 0.9886 |
| $Q_{G_4}$ | 0.4733 | $U_{13}$ | 0.9805 |
| $Q_{G_5}$ | 0.2509 | $U_{14}$ | 0.9441 |
| $U_1$ | 1.0500 | $T_1$ | 0.17 |
| $U_2$ | 1.0266 | $T_2$ | 1.075 |
| $U_3$ | 0.9689 | $T_3$ | 1.025 |
| $U_4$ | 0.9993 | $Q_{C_1}$ | 1.025 |

The detailed comparisons of *PI<sub>c</sub>* are listed in Table 7, and the relative errors are listed in Table 8, with each branch exiting from operation. In addition, the ANN and SVM algorithms are employed for contrast, and the results are also listed in the tables.

**Table 7.** Security index results of each method in the base load condition of the IEEE 14-bus system.

| Outage Line | NRLF | MSA-Lasso Prediction | ANN Prediction | SVM Prediction |
|---|---|---|---|---|
| L 1(1–2) | 0.6467 | 0.6485 | 0.6430 | 0.6392 |
| L 2(1–5) | 0.4065 | 0.4069 | 0.4090 | 0.4046 |
| L 3(2–3) | 0.4184 | 0.4171 | 0.4188 | 0.4163 |
| L 4(2–4) | 0.4798 | 0.4808 | 0.4842 | 0.4843 |
| L 5(2–5) | 0.3506 | 0.3515 | 0.3481 | 0.3519 |
| L 6(3–4) | 0.1470 | 0.1472 | 0.1469 | 0.1475 |
| L 7(4–5) | 0.7802 | 0.7814 | 0.7752 | 0.7766 |
| L 8(4–7) | 0.2765 | 0.2769 | 0.2788 | 0.2790 |
| L 9(4–9) | 0.6334 | 0.6329 | 0.6282 | 0.6393 |
| L 10(5–6) | 0.1586 | 0.1587 | 0.1595 | 0.1585 |
| L 11(6–11) | 2.1960 | 2.1909 | 2.1858 | 2.2013 |
| L 12(6–12) | 0.5327 | 0.5333 | 0.5364 | 0.5329 |
| L 13(6–13) | 2.5532 | 2.5450 | 2.5356 | 2.5377 |
| L 15(7–9) | 2.1749 | 2.1714 | 2.1859 | 2.1687 |
| L 16(9–10) | 0.4327 | 0.4313 | 0.4310 | 0.4304 |
| L 17(9–14) | 1.0499 | 1.0470 | 1.0558 | 1.0458 |
| L 18(10–11) | 1.3361 | 1.3387 | 1.3349 | 1.3322 |
| L 19(12–13) | 0.3816 | 0.3820 | 0.3845 | 0.3818 |
| L 20(13–14) | 2.4478 | 2.4445 | 2.4350 | 2.4317 |

**Table 8.** Relative errors of each method in the base load condition of the IEEE 14-bus system.

| Outage line | Errors of MSA-Lasso (%) | Errors of ANN (%) | Errors of SVM (%) |
|---|---|---|---|
| L 1(1–2) | 0.2909 | −0.5716 | −1.1521 |
| L 2(1–5) | 0.0857 | 0.6023 | −0.4704 |
| L 3(2–3) | −0.3148 | 0.1012 | −0.4888 |
| L 4(2–4) | 0.2033 | 0.8992 | 0.9360 |
| L 5(2–5) | 0.2789 | −0.7009 | 0.3732 |
| L 6(3–4) | 0.0790 | −0.0655 | 0.3152 |
| L 7(4–5) | 0.1494 | −0.6513 | −0.4715 |
| L 8(4–7) | 0.1532 | 0.8395 | 0.9044 |
| L 9(4–9) | −0.0933 | −0.8291 | 0.9291 |
| L 10(5–6) | 0.0641 | 0.5136 | −0.0713 |
| L 11(6–11) | −0.2285 | −0.4608 | 0.2441 |
| L 12(6–12) | 0.1106 | 0.6771 | 0.0276 |
| L 13(6–13) | −0.3212 | −0.6900 | −0.6056 |
| L 15(7–9) | −0.1630 | 0.5034 | −0.2870 |
| L 16(9–10) | −0.3028 | −0.3842 | −0.5136 |
| L 17(9–14) | −0.2799 | 0.5574 | −0.3959 |
| L 18(10–11) | 0.1988 | −0.0852 | −0.2870 |
| L 19(12–13) | 0.1136 | 0.7502 | 0.0576 |
| L 20(13–14) | −0.1336 | −0.5202 | −0.6567 |

From the above two tables, it can be seen that the values of $PI_c$ obtained by algorithm prediction are basically close to those obtained by direct calculation. Compared with ANN and SVM, the MSA-Lasso algorithm has the advantage for predicting the security indices $PI_c$, because the maximum relative error of MSA-Lasso is the smallest among the three algorithms. For a better presentation, the relative errors are shown in Figure 5.

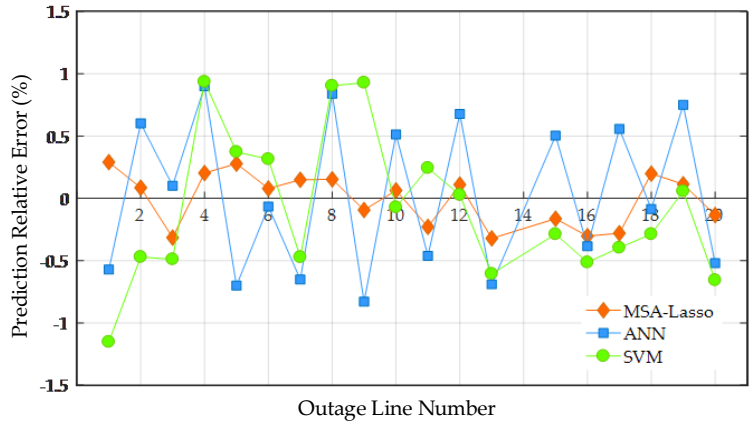

**Figure 5.** Security indices of each method in the base load condition of the IEEE 14-bus system.

As shown in Figure 5, the MSA-Lasso algorithm has a narrow error range of initiation in the three algorithms, which means that the MSA-Lasso algorithm is more stable and accurate than the others. Further discussion of the comparisons is based on the boxplot of relative errors, as shown in Figure 6.

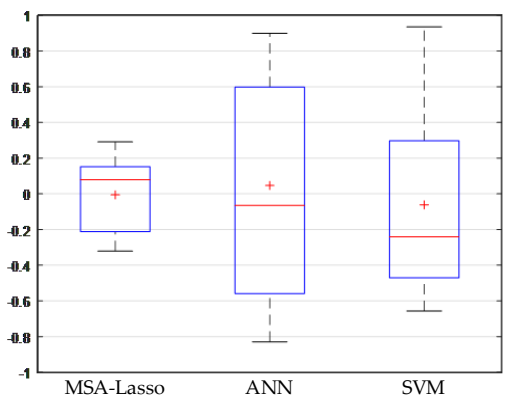

**Figure 6.** Boxplot of relative errors in the base load condition of the IEEE 14-bus system.

In the figure, the red crosses represent the mean of errors, and the red lines are the median (50th percentile) of the errors. From the above figure, it can be seen that although the median of ANN is the best, the mean of MSA-Lasso is closer to zero. Thus, MSA-Lasso has a small deviation in the three algorithms. By comparing the lengths of the boxes that reflect dispersion, it is obvious that MSA-Lasso is the best. Therefore, it can be concluded that the MSA-Lasso algorithm has an advantage in the predictions compared with other algorithms. Then, the security indices are sorted in descending order, and the larger value of $PI_c$ means that the contingency is more serious. The sorted results of all of the lines are listed in Table 9, and the digits represent branch numbers.

**Table 9.** Ranking of security indices in the base load condition of the IEEE 14-bus system.

| Method | Ranking |
|---|---|
| NRLF | L 13  L 20  L 11  L 15  L 18  L 17  L 7  L 1  L 9  L 12  L 4  L 16  L 3  L 2  L 19  L 5  L 8  L 10  L 6 |
| MSA-Lasso | L 13  L 20  L 11  L 15  L 18  L 17  L 7  L 1  L 9  L 12  L 4  L 16  L 3  L 2  L 19  L 5  L 8  L 10  L 6 |
| ANN | L 13  L 20  L 15  L 11  L 18  L 17  L 7  L 1  L 9  L 12  L 4  L 16  L 3  L 2  L 19  L 5  L 8  L 10  L 6 |
| SVM | L 13  L 20  L 11  L 15  L 18  L 17  L 7  L 9  L 1  L 12  L 4  L 16  L 3  L 2  L 19  L 5  L 8  L 10  L 6 |

From the results, it is clear that the ranking of the MSA-Lasso algorithm is the same as that via NRLF. However, the ANN and SVM algorithms have little differences with NRLF. The results show that the MSA-Lasso algorithm is more suitable for the prediction of security indices. The reason why the others have different orders is that the three algorithms are all biased estimation algorithms [33]; the prediction errors may affect the contingency ranking orders when the values of *PIc* are relatively close. The ranking results with the descending order of index values via the NFLR method are shown in Figure 7.

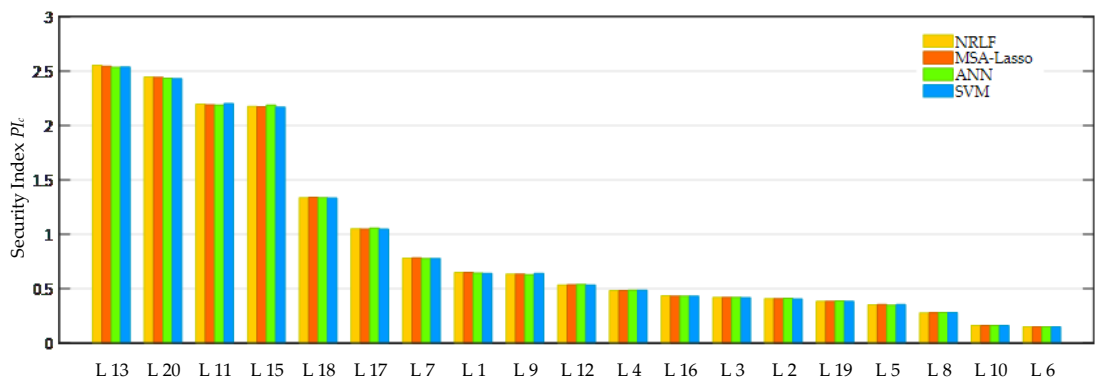

**Figure 7.** Ranking results of each method in the base load condition of the IEEE 14-bus system.

From the above analysis, the proposal has proven that it is effective on the base load condition of the IEEE 14-bus system, since it can predict the security index within a small error, and all of the lines can be identified in a proper order.

5.1.3. Light and Heavy Load Conditions

For testing the accuracy of the prediction of the MSA-Lasso, ANN, and SVM algorithms in the light and heavy load conditions of the IEEE 14-bus system, 80% and 110% of the base load situations are employed for representing light and heavy conditions, respectively. The tested controlled variables are still applied to the variables in Table 6. The ranking results of each condition, with the descending order of the NRLF method, are shown in Tables 10 and 11.

**Table 10.** Security index results of each method in the light load condition of the IEEE 14-bus system.

| Outage Line | NRLF | MSA-Lasso Prediction | ANN Prediction | SVM Prediction |
| --- | --- | --- | --- | --- |
| L 20(13–14) | 1.9557 | 1.9500 | 1.9677 | 1.9786 |
| L 13(6–13) | 1.8806 | 1.8829 | 1.8790 | 1.8698 |
| L 11(6–11) | 1.7781 | 1.7739 | 1.7880 | 1.8013 |
| L 15(7–9) | 1.5623 | 1.5621 | 1.5717 | 1.5622 |
| L 18(10–11) | 0.9984 | 0.9995 | 0.9931 | 0.9977 |
| L 17(9–14) | 0.5178 | 0.5162 | 0.5146 | 0.5190 |
| L 7(4–5) | 0.3878 | 0.3870 | 0.3890 | 0.3899 |
| L 9(4–9) | 0.3014 | 0.3005 | 0.3040 | 0.3018 |
| L 12(6–12) | 0.2111 | 0.2118 | 0.2111 | 0.2102 |
| L 1(1–2) | 0.1550 | 0.1551 | 0.1551 | 0.1564 |
| L 4(2–4) | 0.1448 | 0.1451 | 0.1457 | 0.1453 |
| L 16(9–10) | 0.1335 | 0.1334 | 0.1325 | 0.1330 |
| L 2(1–5) | 0.0985 | 0.0985 | 0.0978 | 0.0978 |
| L 19(12–13) | 0.0832 | 0.0830 | 0.0831 | 0.0843 |
| L 3(2–3) | 0.0748 | 0.0750 | 0.0753 | 0.0749 |
| L 5(2–5) | 0.0417 | 0.0417 | 0.0419 | 0.0417 |
| L 6(3–4) | 0 | 0 | 0 | 0 |
| L 8(4–7) | 0 | 0 | 0.0013 | 0 |
| L 10(5–6) | 0 | 0 | 0 | 0 |

Table 11. Security index results of each method in the heavy load condition of the IEEE 14-bus system.

| Outage Line | NRLF | MSA-Lasso Lrediction | ANN Prediction | SVM Prediction |
| --- | --- | --- | --- | --- |
| L 13(6–13) | 2.9115 | 2.9148 | 2.8915 | 2.9455 |
| L 20(13–14) | 2.7138 | 2.7206 | 2.7142 | 2.7071 |
| L 15(7–9) | 2.5116 | 2.5062 | 2.5020 | 2.4787 |
| L 11(6–11) | 2.4284 | 2.4343 | 2.4261 | 2.4788 |
| L 18(10–11) | 1.5278 | 1.5276 | 1.5397 | 1.5399 |
| L 17(9–14) | 1.3259 | 1.3278 | 1.3154 | 1.3394 |
| L 1(1–2) | 1.0056 | 1.0044 | 1.0017 | 1.0093 |
| L 7(4–5) | 0.9890 | 0.9888 | 0.9845 | 0.9860 |
| L 9(4–9) | 0.8118 | 0.8128 | 0.8092 | 0.8155 |
| L 12(6–12) | 0.7024 | 0.7015 | 0.7027 | 0.7021 |
| L 4(2–4) | 0.6649 | 0.6658 | 0.6624 | 0.6723 |
| L 3(2–3) | 0.6120 | 0.6104 | 0.6132 | 0.6128 |
| L 16(9–10) | 0.5946 | 0.5926 | 0.5979 | 0.6003 |
| L 2(1–5) | 0.5727 | 0.5733 | 0.5720 | 0.5710 |
| L 19(12–13) | 0.5388 | 0.5387 | 0.5412 | 0.5358 |
| L 5(2–5) | 0.5155 | 0.5141 | 0.5166 | 0.5171 |
| L 8(4–7) | 0.4615 | 0.4622 | 0.4585 | 0.4624 |
| L 10(5–6) | 0.3850 | 0.3860 | 0.3839 | 0.3872 |
| L 6(3–4) | 0.3078 | 0.3081 | 0.3091 | 0.3086 |

With the analysis of the light, normal, and heavy load conditions, it can be seen that the rankings of the security indices are basically the same. In addition, with the aggravating of loads, the insecure states will gradually increase.

From the results in this section, the conclusion can be drawn that the MSA-Lasso algorithm is also suitable for the light and heavy load conditions of the IEEE 14-bus system. With the utilization of MSA-Lasso, the severities of the *N*−1 contingencies can be ranked, and the contingencies can be screened.

*5.2. IEEE 118-Bus System*

The IEEE 118-bus system [9,16,20,36], which has 54 generators, 91 loads, 186 branches, nine adjustable transformers, and 12 reactive power compensation devices, is then utilized for examining the effectiveness. A single line diagram of the system is shown in Figure 8.

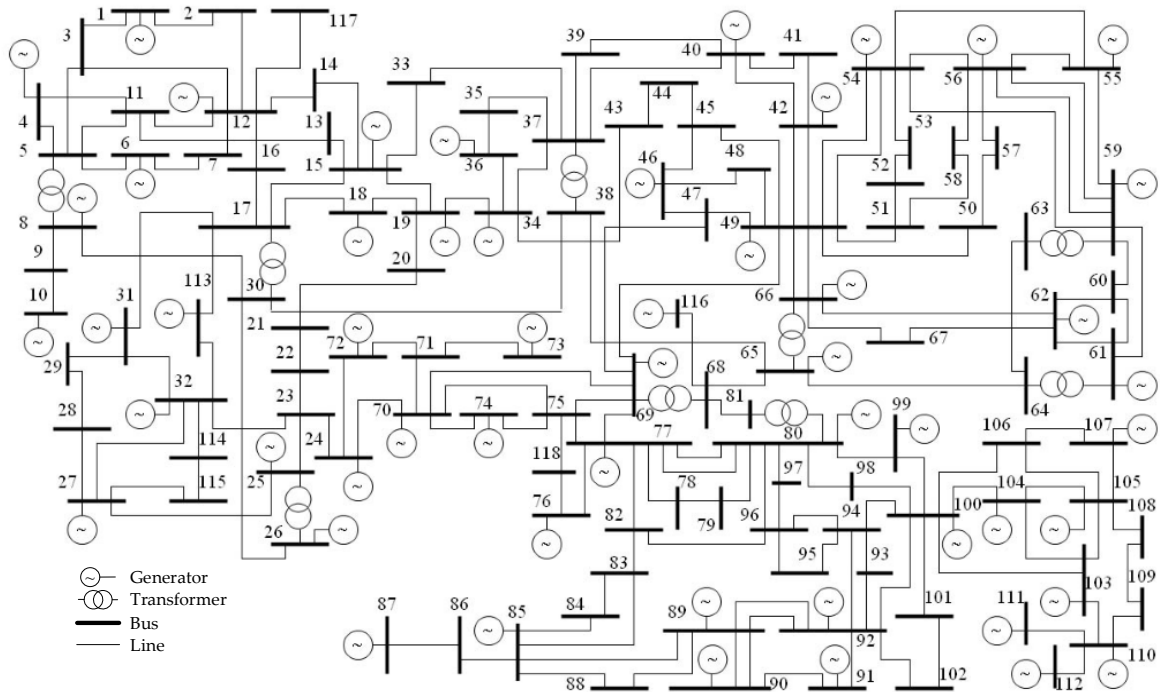

**Figure 8.** Single line diagram of the IEEE 118-bus system.

In the case of the 118-bus system, 179 lines are considered for the *N*−1 contingencies analysis. In addition, 8950 sets are generated for the MSA-Lasso algorithm.

5.2.1. Base Load Conditions

Under the condition of *N*−1 contingencies in the normal load conditions of the IEEE 118-bus system, the values of $PI_c$ are respectively calculated by employing NRLF analysis and the MSA-Lasso algorithm. Some 30 branches account for more than 15% of all lines, and are randomly selected in two results obtained by the two methods [16], which can represent different operating states of the power system. With descending order of $PI_c$ values obtained by NRLF, the outage lines, corresponding $PI_c$, and relative errors are listed in Table 12.

**Table 12.** Security index results of two methods in the base load condition of the IEEE 118-bus system.

| Outage Line | NRLF | MSA-Lasso | Relative Error (%) |
|---|---|---|---|
| L 133(85–86) | 45.3983 | 45.5113 | 0.2490 |
| L 7(8–9) | 5.6776 | 5.6735 | −0.0724 |
| L 97(64–65) | 3.2101 | 3.2149 | 0.1500 |
| L 60(34–43) | 2.6039 | 2.5991 | −0.1851 |
| L 121(77–78) | 1.7891 | 1.7866 | −0.1439 |
| L 104(65–68) | 1.5820 | 1.5841 | 0.1347 |
| L 128(77–82) | 1.3925 | 1.3961 | 0.2607 |
| L 74(53–54) | 1.3513 | 1.3500 | −0.0965 |
| L 40(29–31) | 1.2995 | 1.3023 | 0.2105 |
| L 147(94–95) | 1.2748 | 1.2757 | 0.0679 |
| L 155(94–100) | 1.1396 | 1.1389 | −0.0633 |
| L 160(100–101) | 0.8759 | 0.8780 | 0.2436 |
| L 156(95–96) | 0.8673 | 0.8685 | 0.1333 |
| L 68(45–49) | 0.7681 | 0.7699 | 0.2366 |
| L 103(66–67) | 0.6108 | 0.6108 | −0.0029 |

| Outage Line | | | |
|---|---|---|---|
| L 151(80–97) | 0.5323 | 0.5312 | −0.2069 |
| L 169(105–106) | 0.4454 | 0.4448 | −0.1327 |
| L 35(28–29) | 0.3897 | 0.3905 | 0.2001 |
| L 131(83–85) | 0.3318 | 0.3302 | −0.4858 |
| L 157(96–97) | 0.2903 | 0.2892 | −0.3773 |
| L 135(85–88) | 0.2532 | 0.2528 | −0.1366 |
| L 123(77–80) | 0.2277 | 0.2275 | −0.0820 |
| L 106(49–69) | 0.2091 | 0.2082 | −0.4515 |
| L 114(70–74) | 0.2088 | 0.2088 | 0.0060 |
| L 108(69–70) | 0.2087 | 0.2091 | 0.1831 |
| L 166(103–105) | 0.2073 | 0.2064 | −0.4133 |
| L 164(100–104) | 0.2044 | 0.2041 | −0.1624 |
| L 142(89–92) | 0.1972 | 0.1966 | −0.2829 |
| L 139(89–90) | 0.1887 | 0.1892 | 0.2562 |
| L 141(89–92) | 0.1529 | 0.1529 | −0.0317 |

From the above comparison, it can be seen that the results calculated by NRLF and predicted by the MSA-Lasso algorithm have essentially the same ranking. What's more, the security index value of the MSA-Lasso prediction has little difference with that of direct computation in the same outage line. The different orders appear in the alarm states ($0 < PI_c \leq 1$), and the values of indices are so close that the errors affect the ranking. The differences in alarm operating states will not affect screening contingencies [16]. Therefore, the MSA-Lasso algorithm is still effective in the IEEE 118-bus system.

The computation time of the proposed OSSA module is then compared with that of NRLF and that from the literature [16] by using a multi-layer feed forward network (MLFFN) and radial basis function network (RBFN). From the table, it is clear that MSA-Lasso is more rapid in the OSSA application.

**Table 13.** Comparison of computation times of the 118-bus system. MLFFN: multi-layer feed forward network; RBFN: radial basis function network.

| Method | NRLF | MSA-Lasso | MLFFN [16] | RBFN [16] |
|---|---|---|---|---|
| Time (s) | 12.269 | 0.875 | 1.438 | 1.172 |

From the above analysis, the Lasso module is suitable for online application, and the effectiveness of the proposed method for 118-bus systems is verified.

5.2.2. Light and Heavy Load Conditions

For testing the effectiveness of the proposed method in light and heavy load conditions of the IEEE 118-bus system, 80% and 110% of the base load situations are respectively employed. The tested controlled variables are the same as those in Section 5.2.1. The ranking results of each condition in descending order are shown in Table 14.

**Table 14.** Security index results of light and heavy load conditions of the IEEE 118-bus system.

| Outage Line | Light Load Condition | Heavy Load Condition |
|---|---|---|
| L 133(85–86) | 44.3598 | 45.8894 |
| L 7(8–9) | 5.6771 | 5.6818 |
| L 97(64–65) | 3.2004 | 3.2143 |
| L 60(34–43) | 1.9922 | 2.9717 |
| L 121(77–78) | 1.7229 | 1.9869 |
| L 104(65–68) | 1.5353 | 1.7278 |
| L 128(77–82) | 1.3043 | 1.6184 |
| L 74(53–54) | 1.2932 | 1.5375 |
| L 40(29–31) | 1.1212 | 1.3955 |
| L 147(94–95) | 1.0863 | 1.2815 |
| L 155(94–100) | 0.9742 | 1.1406 |
| L 160(100–101) | 0.8713 | 0.9323 |
| L 156(95–96) | 0.8360 | 0.8709 |
| L 68(45–49) | 0.7522 | 0.0835 |
| L 103(66–67) | 0.5986 | 0.6553 |
| L 151(80–97) | 0.5215 | 0.5474 |
| L 169(105–106) | 0.4339 | 0.5734 |
| L 35(28–29) | 0.4283 | 0.4032 |
| L 131(83–85) | 0.4186 | 0.3385 |
| L 157(96–97) | 0.2857 | 0.3057 |
| L 135(85–88) | 0.2861 | 0.2875 |
| L 123(77–80) | 0.2069 | 0.2736 |
| L 106(49–69) | 0.2038 | 0.2687 |
| L 114(70–74) | 0.2022 | 0.2676 |
| L 108(69–70) | 0.2012 | 0.2682 |
| L 166(103–105) | 0.2004 | 0.2669 |
| L 164(100–104) | 0.1983 | 0.2656 |
| L 142(89–92) | 0.1928 | 0.2335 |
| L 139(89–90) | 0.1690 | 0.1997 |
| L 141(89–92) | 0.1385 | 0.1538 |

From the data of Table 12 and the above table in the IEEE 118-bus system, it can be seen that the rankings of the outage lines are basically the same. Moreover, for the same outage line, the heavier loads cause a large security index value, and the conclusion can be drawn that the proposal is suitable for light, normal, and heavy load conditions in the larger scale system.

*5.3. IEEE 300-bus System*

For examining the effectiveness of large-scale power systems, the IEEE 300-bus system is applied as an example in this paper, which has 69 generators, 68 loads, 411 branches, 107 adjustable transformers, and 14 reactive power compensation devices [18]; the system is available in [36].

In this case, 342 lines are considered, and 17,100 sets are generated for the MSA-Lasso algorithm.

5.3.1. Base Load Conditions

Under the condition of $N-1$ contingencies in the IEEE 300-bus system, the values of security index $PI_c$ are computed by using two methods, which are respectively calculation by NPLF and prediction via the MSA-Lasso algorithm. The time for contingency screening and ranking by adopting the proposed module is 4.1732 s, while the computation time is 76.8578 s by using NPLF. Obviously, the proposal is available for online application. Some 60 branches accounting for more than 15% of all of the lines are randomly selected, which can represent different operating states of the power system. The results of $PI_c$ by adopting two ways, and the relative errors, are listed in Table 15 in the descending order of the NPLF method.

**Table 15.** Security index results of two methods in the base load condition of the IEEE 300-bus system.

| Outage Line | NRLF | MSA-Lasso | Relative Error (%) |
|---|---|---|---|
| L 181(119–120) | 284.1077 | 283.5923 | −0.1814 |
| L 309(225–191) | 50.4040 | 50.5872 | 0.3635 |
| L 114(59–61) | 19.5083 | 19.5364 | 0.1436 |
| L 242(162–164) | 17.2150 | 17.2048 | −0.0592 |
| L 257(178–180) | 6.7417 | 6.7493 | 0.1133 |
| L 322(241–237) | 6.7152 | 6.7214 | 0.0917 |
| L 246(167–169) | 3.1995 | 3.2130 | 0.4206 |
| L 205(133–137) | 1.2821 | 1.2851 | 0.2337 |
| L 10(9006–9007) | 1.1665 | 1.1688 | 0.1944 |
| L 59(16–42) | 1.1353 | 1.1391 | 0.3343 |
| L 249(173–174) | 1.0834 | 1.0789 | −0.4179 |
| L 210(134–184) | 1.0549 | 1.0556 | 0.0670 |
| L 174(115–122) | 0.8585 | 0.8621 | 0.4129 |
| L 273(194–664) | 0.4088 | 0.4089 | 0.0342 |
| L 207(133–169) | 0.3709 | 0.3695 | −0.3729 |
| L 308(224–226) | 0.3678 | 0.3681 | 0.0611 |
| L 275(196–197) | 0.1832 | 0.1828 | −0.2299 |
| L 86(38–41) | 0.1707 | 0.1707 | −0.0462 |
| L 323(240–281) | 0.1479 | 0.1474 | −0.3416 |
| L 140(81–194) | 0.1469 | 0.1465 | −0.3172 |
| L 266(190–231) | 0.1356 | 0.1359 | 0.2660 |
| L 85(37–90) | 0.1340 | 0.1337 | −0.2068 |
| L 143(86–87) | 0.1338 | 0.1338 | 0.0051 |
| L 9(9005–9055) | 0.1330 | 0.1330 | 0.0251 |
| L 142(85–86) | 0.1276 | 0.1270 | −0.4102 |
| L 274(195–219) | 0.1240 | 0.1244 | 0.3311 |
| L 192(136–158) | 0.1217 | 0.1217 | −0.0293 |
| L 227(143–145) | 0.1206 | 0.1211 | 0.4201 |
| L 185(123–124) | 0.0989 | 0.0990 | 0.0652 |
| L 159(99–109) | 0.0881 | 0.0884 | 0.3287 |
| L 197(128–130) | 0.0819 | 0.0817 | −0.2183 |
| L 184(122–125) | 0.0783 | 0.0780 | −0.3255 |
| L 163(103–105) | 0.0782 | 0.0780 | −0.2962 |
| L 162(102–104) | 0.0737 | 0.0734 | −0.3757 |
| L 239(157–159) | 0.0726 | 0.0724 | −0.2540 |
| L 324(242–245) | 0.0671 | 0.0673 | 0.3236 |
| L 128(73–79) | 0.0663 | 0.0665 | 0.2394 |
| L 54(13–20) | 0.0654 | 0.0656 | 0.3786 |
| L 112(57–63) | 0.0645 | 0.0643 | −0.3573 |
| L 251(173–176) | 0.0641 | 0.0639 | −0.3667 |
| L 201(130–132) | 0.0632 | 0.0634 | 0.4178 |
| L 65(22–23) | 0.0619 | 0.0620 | 0.0239 |
| L 314(228–234) | 0.0614 | 0.0616 | 0.2539 |
| L 248(172–174) | 0.0613 | 0.0612 | −0.1593 |
| L 215(137–181) | 0.0611 | 0.0610 | −0.3247 |
| L 256(178–179) | 0.0610 | 0.0611 | −0.0074 |
| L 295(214–242) | 0.0608 | 0.0608 | −0.0681 |
| L 315(229–190) | 0.0604 | 0.0606 | 0.3647 |
| L 340(10–11) | 0.0599 | 0.0598 | −0.2893 |
| L 221(140–145) | 0.0596 | 0.0594 | −0.2511 |
| L 120(69–79) | 0.0593 | 0.0591 | −0.2974 |
| L 222(140–146) | 0.0588 | 0.0591 | 0.4009 |
| L 14(9012–9002) | 0.0526 | 0.0527 | 0.2202 |
| L 335(3–1) | 0.0477 | 0.0476 | −0.2486 |
| L 48(7–131) | 0.0466 | 0.0467 | 0.1484 |
| L 80(37–38) | 0.0322 | 0.0322 | −0.2373 |
| L 7(9005–9053) | 0.0289 | 0.0288 | −0.0387 |
| L 21(9007–9071) | 0.0073 | 0.0074 | 0.3859 |
| L 34(9003–9036) | 0.0073 | 0.0073 | 0.0760 |
| L 118(63–526) | 0.0052 | 0.0052 | −0.2994 |

From the above table, it can be seen that the values of $PI_c$ obtained by adopting the two methods have similar ranking results, and the contingencies can also be screened. Thus, the conclusion can be drawn that the proposal is also effective for large-scale power systems.

5.3.2. Light and Heavy Load Conditions

For testing the effectiveness of the proposed method in light and heavy load conditions of the IEEE 300-bus system, the conditions of 80% and 110% of base load are respectively applied. The tested controlled variables are the same as those in Section 5.3.1. The ranking results of light and heavy conditions in descending order are shown in Table 16.

Table 16. Security index results of light and heavy load conditions of the IEEE 300-bus system.

| Outage Line | Light Load Condition | Heavy Load Condition |
|---|---|---|
| L 181(119–120) | 147.7148 | 341.7919 |
| L 309(225–191) | 40.2870 | 70.5317 |
| L 114(59–61) | 23.7148 | 39.3177 |
| L 242(162–164) | 13.3284 | 35.5648 |
| L 257(178–180) | 7.2198 | 11.5840 |
| L 322(241–237) | 6.4276 | 11.2255 |
| L 246(167–169) | 1.4334 | 9.0839 |
| L 205(133–137) | 0.9297 | 9.9235 |
| L 10(9006–9007) | 0.9201 | 3.7961 |
| L 59(16–42) | 0.8912 | 3.4846 |
| L 249(173–174) | 0.8763 | 3.2126 |
| L 210(134–184) | 0.7992 | 3.0168 |
| L 174(115–122) | 0.7311 | 2.8433 |
| L 273(194–664) | 0.3252 | 2.7788 |
| L 207(133–169) | 0.3732 | 2.5910 |
| L 308(224–226) | 0.2521 | 2.3859 |
| L 275(196–197) | 0.1436 | 2.0399 |
| L 86(38–41) | 0.1418 | 1.9632 |
| L 323(240–281) | 0.1586 | 1.9495 |
| L 140(81–194) | 0.1513 | 1.9282 |
| L 266(190–231) | 0.1151 | 1.9248 |
| L 85(37–90) | 0.1146 | 1.8892 |
| L 143(86–87) | 0.0945 | 1.8849 |
| L 9(9005–9055) | 0.0944 | 1.8741 |
| L 142(85–86) | 0.0841 | 1.7082 |
| L 274(195–219) | 0.0834 | 1.6927 |
| L 192(136–158) | 0.0834 | 1.6454 |
| L 227(143–145) | 0.0825 | 1.6394 |
| L 185(123–124) | 0.0525 | 1.9374 |
| L 159(99–109) | 0.0524 | 1.8280 |
| L 197(128–130) | 0.0523 | 1.7938 |
| L 184(122–125) | 0.0515 | 1.7828 |
| L 163(103–105) | 0.0418 | 1.5820 |
| L 162(102–104) | 0.0491 | 1.5415 |
| L 239(157–159) | 0.0507 | 1.5416 |
| L 324(242–245) | 0.0476 | 1.5606 |
| L 128(73–79) | 0.0476 | 1.5222 |
| L 54(13–20) | 0.0477 | 1.4659 |
| L 112(57–63) | 0.0467 | 1.5399 |
| L 251(173–176) | 0.0466 | 1.5381 |
| L 201(130–132) | 0.0464 | 1.4931 |
| L 65(22–23) | 0.0464 | 1.3600 |
| L 314(228–234) | 0.0464 | 1.2715 |
| L 248(172–174) | 0.0455 | 1.2575 |
| L 215(137–181) | 0.0492 | 1.1494 |
| L 256(178–179) | 0.0474 | 1.1425 |
| L 295(214–242) | 0.0460 | 1.0201 |
| L 315(229–190) | 0.0388 | 1.0044 |
| L 340(10–11) | 0.0290 | 0.9898 |
| L 221(140–145) | 0.0249 | 0.9442 |
| L 120(69–79) | 0.0243 | 0.9717 |
| L 222(140–146) | 0.0181 | 0.9353 |
| L 14(9012–9002) | 0.0155 | 0.8952 |
| L 335(3–1) | 0.0134 | 0.8542 |
| L 48(7–131) | 0.0125 | 0.8440 |
| L 80(37–38) | 0.0100 | 0.0762 |
| L 7(9005–9053) | 0.0084 | 0.0761 |
| L 21(9007–9071) | 0.0062 | 0.0378 |
| L 34(9003–9036) | 0 | 0.0092 |
| L 118(63–526) | 0 | 0.0076 |

From the values of the security indices in Table 16, a similar conclusion can be drawn that the rankings are almost unchanged, and increasing the loads will lead to serious consequences. What's more, the effectiveness of the proposal in the IEEE 300-bus system is verified.

## 6. Conclusions

In terms of fast and accurate contingency screening and ranking, an online static security assessment module based on a multi-step adaptive Lasso regression algorithm is proposed in this paper. The proposed approach is examined on the IEEE 14-bus, 118-bus, and 300-bus test systems, and the results indicate that this approach manages to handle this issue with reduced time, and is suitable for online application. The following conclusions can be drawn from the work:

(1) Based on the online static security assessment module in this paper, the issues, which include operating state identifying and contingency screening and ranking, can be solved quickly and accurately. What's more, the operating state considers the impacts of transformers and compensation devices, and subsequently realizes better control of power systems.
(2) Due to the proposed method not needing to calculate a large number of load flow under the conditions of contingencies and the MSA-Lasso algorithm having more accuracy than the other learning algorithm, it is suitable for an online assessment of the static security of power systems.
(3) Considering the current various operating states of power systems, the proposed method analyzed different load conditions that varied from 50% to 150% of the base load. Through online static security assessment modules in different load conditions, the MSA-Lasso algorithm can assess the static security problems in the normal, light, and heavy load conditions.

Future research will focus on considering parallel computation techniques in machine learning methods [12]. It will be interesting to develop an efficient version to improve the practicality of the presented approach for real-world applications. Moreover, with the increment of renewable energy and energy storage embedded into the power system [37,38,39], the research of static security assessment considering renewable energy and load uncertainty [40] and integrated energy systems [41] is a meaningful topic.

**Author Contributions:** Conceptualization, Yang Li; Methodology, Yang Li and Yahui Li; Software, Yahui Li; Validation, Yang Li, Yahui Li and Yuanyuan Sun; Formal Analysis, Yang Li and Yahui Li; Investigation, Yang Li and Yahui Li; Resources, Yang Li; Data Curation, Yang Li; Writing-Original Draft Preparation, Yahui Li; Writing-Review & Editing, Yang Li; Visualization, Yang Li and Yahui Li; Supervision, Yang Li and Yuanyuan Sun; Project Administration, Yang Li; Funding Acquisition, Yang Li.

**Funding:** This research was funded by the China Scholarship Council (CSC) grant number 201608220144, and the National Natural Science Foundation of China grant number 51677023.

**Conflicts of Interest:** The authors declare no conflict of interest.

## Abbreviations

| | |
|---|---|
| OSSA | Online static security assessment |
| SSA | Static security assessment |
| DSA | Dynamic security assessment |
| NRLF | Newton–Raphson load flow |
| Lasso | Least absolute shrinkage and selection operator |
| MSA-Lasso | Multi-step adaptive least absolute shrinkage and selection operator |
| ANN | Artificial neural network |
| SVM | Support vector machine |
| **Y** | The vector consisting of responses |
| **X** | The input matrix that is formed by observations |
| $S$ | The size of the training set |
| $D$ | The number of controlled variables |
| $\lambda$ | The shrinkage tuning parameter |
| $P_G$ | Active power output of generator |
| $Q_G$ | Reactive power output of generator |
| $N_G$ | The number of generators |
| $P_D$ | Active load of load bus |
| $Q_D$ | Reactive load of load bus |
| $U$ | Voltage amplitude of bus |
| $\delta$ | Voltage angle of bus |
| $N_{ac}$ | The numbers of buses |
| $T$ | The tap of transformer |
| $N_T$ | The numbers of adjustable transformer taps |
| $Q_C$ | The switching capacity of reactive power compensation capacitor |
| $N_C$ | The numbers of reactive power compensation capacitor banks |
| $PI_c$ | The composite security index |